%% file: acl_latex.tex
\documentclass[11pt]{article}
\usepackage{acl}

\usepackage{times}
\usepackage{latexsym}
\usepackage{amsmath}   
\usepackage{amssymb}   
\usepackage{booktabs}  
\usepackage{enumitem}
\usepackage[T1]{fontenc}
\usepackage{float}
\usepackage[utf8]{inputenc}

\usepackage{microtype}

\usepackage{inconsolata}

\usepackage{graphicx}
\usepackage{multirow}
\usepackage{subcaption}

\title{AutothinkRAG: Complexity-Aware Control of Retrieval-Augmented Reasoning for Image-Text Interaction}

\author{
  Jiashu Yang$^{1,3}$, 
  Chi Zhang$^{1}$, 
  Abudukelimu Wuerkaixi$^{2,3}$, 
  Xuxin Cheng$^{3}$, \\
  {\bf Cao Liu$^{3}$, 
  Ke Zeng$^{3}$, 
  Xu Jia$^{1}$, 
  Xunliang Cai$^{3}$} \\
  $^1$Dalian University of Technology   $^2$Tsinghua University   $^3$Meituan LongCat Interaction Team \\
}

\begin{document}
\maketitle
\input{sections/abstract}
\input{sections/intro}
\input{sections/related}
\input{sections/method}

\input{sections/experiment}
\input{sections/conclusion}

\bibliography{custom}

\end{document}

%% file: sections/abstract.tex

\begin{abstract}
Multimodal document question answering requires retrieving dispersed evidence from visually rich long documents and performing reliable reasoning over heterogeneous information. Existing multimodal RAG systems remain limited by two bottlenecks: static retrieval that ignores query complexity, and end-to-end Vision-Language Models (VLMs) that couple visual perception with logical reasoning, leading to inefficient computation and unstable answer generation. We propose AutoThinkRAG, a complexity-aware inference architecture for multimodal document QA. It has two components: (1) a Query Complexity Router that analyzes query difficulty and structure to adaptively select retrieval and reasoning paths; and (2) a Perception--Reasoning Decoupling architecture that uses a lightweight VLM as a high-fidelity visual interpreter to convert query-relevant visual cues into textual representations, which are then passed to an LLM for logical reasoning and answer synthesis. This design improves both efficiency and robustness, especially on long-document and unanswerable queries. Experiments on DocBench and MMLongBench show that AutoThinkRAG achieves 82.13\% and 51.29\% overall accuracy, respectively, while reducing per-query token consumption by 18.9\% and monetary cost by 18.2\%. Further analyses show that the gains are most pronounced on complex queries requiring adaptive retrieval and multi-step reasoning.
\end{abstract}

%% file: sections/intro.tex
\section{Introduction}
\input{image/motis}

Retrieval-Augmented Generation (RAG) serves as a core technology for processing heterogeneous multimodal documents. Its objective is to retrieve multimodal information relevant to a given query from documents as contextual content for question answering, thereby enhancing the accuracy and richness of generated results. As an evolving direction in this field, Multimodal RAG (MM-RAG) needs to address two core issues simultaneously: multimodal information processing and high-precision reasoning.

Knowledge graphs can effectively represent the relationships and attributes of information, while dense vector representations can identify semantically relevant content lacking direct structural connections. In recent years, researchers have conducted extensive explorations on multimodal graph RAG and complex question answering. As the state-of-the-art (SOTA) approach, RAG-Anything \citep{guo2025raganything} has validated the application of multimodal graphs in document question answering. HyperGraphRAG \citep{luo2025hypergraphrag} effectively captures high-order dependencies in complex knowledge bases through hyperedge modeling, improving the accuracy of complex multi-hop reasoning.

Despite the progress achieved by existing methods, current SOTA frameworks still face two critical issues that severely impact performance: (1) \textbf{Retrieval Rigidity}: Existing systems adopt a static retrieval strategy that fails to make reasonable judgments and planning based on query complexity. Accurate understanding of queries with arbitrary complexity requires large-scale models, and this high-computation-cost approach leads to inefficient allocation of computing resources. (2) \textbf{Reasoning Deficit}: Multimodal information question answering relies on monolithic Vision-Language Models (VLMs) for end-to-end generation. However, recent studies have shown that VLMs exhibit significantly lower reasoning performance than Large Language Models (LLMs), leading to the problem of ``correct visual recognition but incorrect answer generation'' \citep{mmgraphrag2024}.

In this paper, we propose the AutothinkRAG framework to address the aforementioned issues. To tackle retrieval rigidity, we design the Autothink lightweight cognitive router, which dynamically selects retrieval paths based on query parsing and complexity evaluation, enabling on-demand allocation of computing resources. To resolve reasoning deficit, we propose the Decomposition of Perception and Reasoning (DPR) architecture, which limits the role of VLMs to that of ``visual descriptors'' and delegates logical integration tasks to LLMs, achieving more accurate reasoning without loss of relevant information.

Our contributions are four-fold: 
\begin{itemize}[itemsep=2pt,topsep=0pt,parsep=0pt]

\item We present AutothinkRAG, a scalable architecture that integrates MinerU-based parsing with a hybrid Graph-Vector storage, establishing a new pareto-optimal frontier between efficiency and accuracy; 
\item We design the Autothink Router to handle queries of unknown complexity; by leveraging a lightweight Small Language Model (SLM) for in-depth complexity analysis and task decomposition, this module enables adaptive execution path selection, effectively addressing the challenge of Retrieval Rigidity.
\item We introduce a decoupled paradigm for multimodal problem solving, which explicitly separates information transformation from the reasoning phase. This approach addresses the limitations of traditional methods that rely on end-to-end direct inference using VLMs.
\item Experiments on two document-understanding benchmark display that AutothinkRAG achieves the new state-of-the-art performance without relying on large-scale models.
\end{itemize}

%% file: image/motis.tex
\begin{figure}[!t]
    \centering
    \includegraphics[width=1.0\columnwidth, clip]{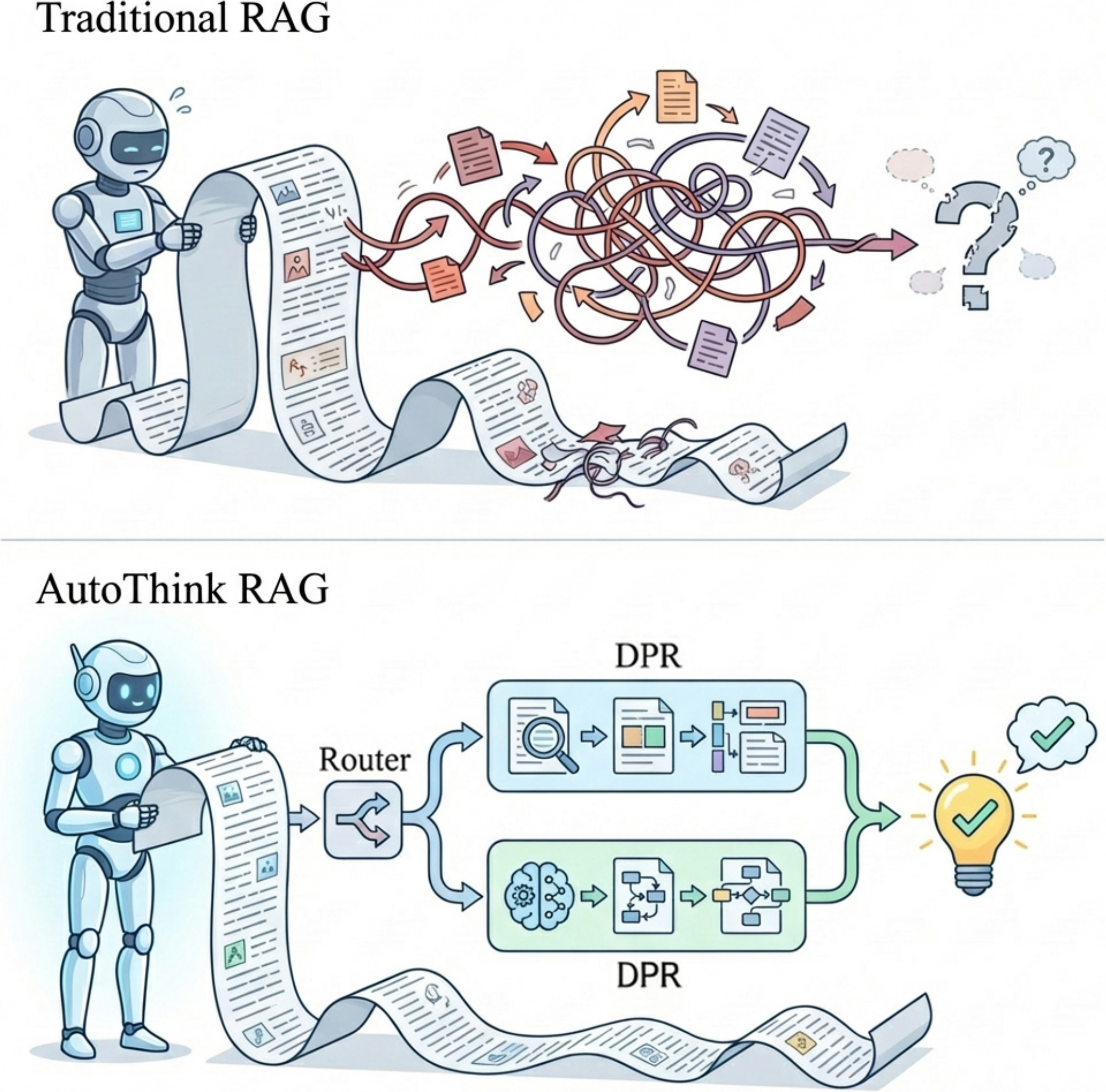}
    \caption{Compared with traditional RAG, AutothinkRAG enables more effective and robust long-document question answering through adaptive retrieval and decoupled perception-and-reasoning.}
    \label{fig:moti}
\end{figure}

%% file: sections/related.tex
\section{Related Work}
\subsection{Graph-based Retrieval-Augmented Generation}
Traditional retrieval-augmented generation (RAG) relies on chunking and vector retrieval, yet often falls short in dispersed evidence localization and multi-hop reasoning over long and multi-document contexts. GraphRAG~\citep{edge2024graphrag} addresses this by modeling relations among retrieval units as graphs to enable structured retrieval and reasoning. Subsequent work improves (i) efficiency and evidence coverage via lightweight/ephemeral graphs and multi-stage retrieval (LightRAG~\citep{guo2024lightrag}, REANO~\citep{li2024reano}, HopRAG~\citep{wang2024hoprag}) and (ii) multi-hop evidence organization by extracting candidate subgraphs/paths and forming structured evidence chains (GNN-RAG~\citep{zhang2024gnnrag}, TRACE the Evidence~\citep{trivedi2023trace}, PathRAG~\citep{liu2024pathrag}). Hierarchical designs further support multi-granularity reasoning (ArchRAG~\citep{wu2024archrag}, From Local to Global~\citep{chen2024localglobal}), while robustness is strengthened through subgraph pruning, subgraph-level retrieval, or external graph knowledge (G-Retriever~\citep{sun2024gretriever}, RAGRAPH~\citep{ma2024ragraph}). Beyond binary relations, hypergraph and higher-arity modeling improves expressiveness and reduces information loss in complex reasoning (HyperGraphRAG~\citep{luo2025hypergraphrag}, Structure Is All You Need~\citep{zhou2024structure}).

\input{image/pipelines}
\subsection{Multimodal Retrieval-Augmented Generation}
Multimodal RAG augments text-based RAG with images and videos as external knowledge sources, mitigating limitations of parametric memory (RA-TTA~\citep{li2023ratta}, MMed-RAG~\citep{zhang2023mmedrag}); however, many systems still depend on textification or single-modality alignment, hindering the preservation of fine-grained cross-modal structures. For visually rich documents, full-page visual retrieval units avoid OCR/linearization loss (VisDOM~\citep{zhao2025visdom}, VisRAG~\citep{singh2024visrag}, ColPali~\citep{delatorre2024colpali}). In image captioning and VQA, robustness issues (e.g., retrieval instability and frequency bias) are commonly mitigated via diversified retrieval and late-interaction mechanisms. For long videos, segment-level or query-aware retrieval reduces computational cost (VideoRAG~\citep{shi2024videorag}). Retrieval also benefits vision--language contrastive learning and generation for improved generalization and alignment (retrieval-enhanced VLMs~\citep{retrievalcontrastivevlm2023}, GarmentAligner~\citep{yang2024garmentaligner}); in domains such as healthcare and content moderation, it introduces high-confidence or personalized priors at the expense of added system complexity~\citep{domainrag2023}.

\subsection{Document Understanding with Retrieval-Augmented Generation}
Long- and multi-document QA is constrained by context length and evidence localization difficulty. Recent methods improve retrieval granularity and context packaging through logical block decomposition and iterative retrieval (Document Haystacks~\citep{chen2025haystacks}), document restructuring and adaptive merging (Virtual Documents~\citep{ma2024virtualdocs}, Document Segmentation Matters~\citep{park2025segmentation}), and structured indexing or hierarchical summarization (PdfTriage~\citep{pdftriage2023}, MemSum-DQA~\citep{memsumdqa2022}, BookRAG~\citep{bookrag2023}). Classical paradigms establish coarse-to-fine localization (Dynamic Coattention Networks~\citep{seo2017dynamic}, Coarse-to-Fine QA~\citep{choi2017coarse}), and later work models global structure and discourse hierarchies (Compressive Graph Selector Networks~\citep{compressivegraphselector2022}, Beyond Chunking~\citep{beyondchunking2023}); for cross-document reasoning, link-aware pretraining further strengthens inter-document semantic modeling~\citep{yasunaga2022linkbert}. Retrieval and generation are increasingly coupled via iterative/corrective schemes (Generate-then-Ground~\citep{zhao2023generate}, REAR~\citep{li2024rear}, Active RAG~\citep{activerag2023}, Iterative Retrieval--Generation Synergy~\citep{iterativerag2023}, IAG~\citep{iag2023}). For visually rich documents, multimodal document RAG incorporates layout, entities, and knowledge structures for multi-page understanding (REVEAL~\citep{hu2023reveal}, VDocRAG~\citep{vdocgrag2023}, VisDoM~\citep{zhao2025visdom}, MMGraphRAG~\citep{mmgraphrag2024}); agent-based methods further improve robustness via multi-agent collaboration and iterative feedback (DocAgent~\citep{docagent2024}, DocLens~\citep{doclens2025}, Doc-React~\citep{docreact2024}).

%% file: image/pipelines.tex
\begin{figure*}[t]
    \centering
    \includegraphics[width=\textwidth]{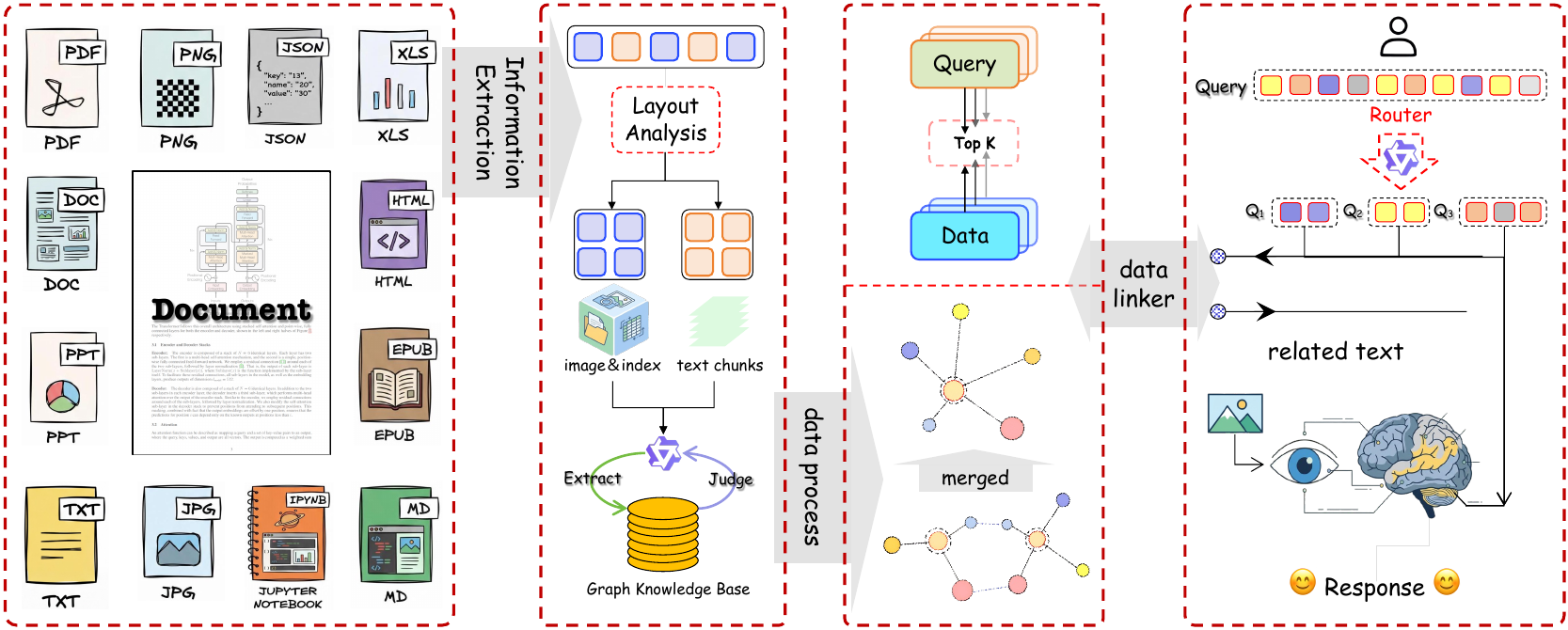}
    \vspace{+0.02in}
    \caption{
    {\textbf{Overview of our AutothinkRAG.} The framework consists of four primary stages: 
    (1) \textbf{Information Extraction}: Heterogeneous multimodal data are ingested. 
    (2) \textbf{Data Processing}: Through layout analysis, the system partitions data into text chunks and image indices, followed by an iterative ``Extract--Judge'' loop to construct a Graph Knowledge Base. 
    (3) \textbf{Data Linker}: A hybrid retrieval mechanism integrates traditional Top--K vector search with graph--based merging to capture complex entity relationships. 
    (4) \textbf{Reasoning \& Response}: The Query Router decomposes the original query into sub-queries ($Q_1, Q_2, Q_3$), which are then processed through a decoupled perception reasoning architecture leveraging both ``related text'' and visual cues to synthesize the final response.}
    }
    \label{fig:pipeline}
\end{figure*}

%% file: sections/method.tex
\section{Methodology}
\label{sec:method}

In this section, we first present the overall workflow of AutothinkRAG. Subsequently, we specify the strategies for information extraction, storage, and transmission. Finally, we detail the implementation of the Query Complexity Router and the Functional Decoupling Architecture.

\subsection{Overall Workflow}
\label{sec:overall_workflow}

AutothinkRAG deconstructs the multimodal Document Question Answering (DocQA) task into intent-aware routing and decoupled reasoning,as illustrated in Figure ~\ref{fig:pipeline}. The end-to-end inference process is formally defined as:
\begin{equation}
A = \Psi\left( Q, \mathcal{D} \right) = \text{LLM}\left( T_v, \mathcal{R}, I_p \right)
\end{equation}
where $Q$ is the input query, $\mathcal{D}$ is the document collection, $\mathcal{R}$ denotes retrieved relevant context, $T_v$ is the textualized visual evidence, $I_p$ is the reasoning path instruction, and $A$ is the final synthesized answer. The workflow proceeds through three core stages:

\begin{enumerate}
    \item \textbf{Knowledge Base Construction}: The system parses $\mathcal{D}$ into content blocks $b$ with metadata $\mathcal{M}_b = \{T, C, B, P, S\}$, encompassing type, content, spatial coordinates, page, and storage path. These are integrated into a hybrid Graph Knowledge Base (GKB) and vector store to support spatial-semantic retrieval (detailed in Sec.~\ref{sec:extraction_storage}).

    \item \textbf{Query Complexity Routing (QCR)}: For each query $Q$, the QCR performs a pre-execution analysis to output a complexity label $c(Q) \in \{\text{Simple, Moderate, Complex}\}$ and routing instructions $I_p$. This stage determines sub-query decomposition and optimizes retrieval paths for the recall module $\mathcal{R}$ (detailed in Sec.~\ref{sec:router_fda}.1).

    \item \textbf{Decomposition of Perception and Reasoning (DPR)}: Guided by $I_p$, the system fetches context $\mathcal{R}$ and invokes raw visual assets via path $S$. A training-free small-scale VLM serves as a visual interpreter to generate structured textual descriptions $T_v$. Finally, a high-performance LLM integrates $\{T_v, \mathcal{R}\}$ to synthesize the final answer $A$ through rigorous deduction following $I_p$ (detailed in Sec.~\ref{sec:router_fda}.2).
\end{enumerate}

By functionally decoupling perception from reasoning, AutothinkRAG bridges the logical bottlenecks in standalone models, ensuring robust comprehension across extensive multimodal contexts.

\subsection{Information Extraction, Storage, and Transmission}
\label{sec:extraction_storage}

To support complex multimodal reasoning, we design a high-fidelity information processing pipeline. This pipeline is engineered to transform heterogeneous documents (e.g., PDF, PPT) into a structured and queryable knowledge system while preserving the original spatial layout and semantic integrity.

\subsubsection{High-Fidelity Parsing and Metadata Extraction}
We employ a robust parsing engine (e.g., MinerU) to perform deep analysis on the input document $D$. The parsing output comprises a structured JSON file containing decomposed content blocks $\mathcal{B}$ and an asset repository for multimodal objects (images, tables, equations). For each extracted block $b \in \mathcal{B}$, we maintain a metadata tuple $\mathcal{M}_b = \{T, C, B, P, S\}$, where:
\begin{itemize}
    \item $T$ denotes the \textbf{Type} (e.g., text, image, or table);
    \item $C$ represents the raw \textbf{Content} or the caption associated with the block;
    \item $B$ is the \textbf{Bounding Box} (Bbox) coordinates for spatial localization;
    \item $P$ indicates the \textbf{Page Number};
    \item $S$ is the \textbf{Storage Path}, serving as the unique index for asset transmission.
\end{itemize}

\subsubsection{Differentiated Information Processing}
The system implements bifurcated semantic extraction logic based on content type:
\begin{itemize}
    \item \textbf{Textual Stream}: Raw text is serialized and partitioned into chunks using a sliding window of 1,200 tokens with a 100-token overlap. Subsequently, an Information Extraction (IE) module identifies entity and relation triplets from each chunk.
    \item \textbf{Multimodal Stream}: For non-textual assets, the system generates a detailed semantic \textbf{Description}. The original \textbf{Caption} is concatenated with this description to form an augmented representation, followed by IE to align visual concepts with the global knowledge schema.
\end{itemize}

\subsubsection{Hybrid Storage and Transmission Protocol}
Processed information is integrated into a dual-layered storage architecture:
\begin{enumerate}
    \item \textbf{Graph Knowledge Base (GKB)}: We employ a hard-matching algorithm for entity disambiguation and merging to construct a global relational graph.
    \item \textbf{Vector Store}: All entities, relations, and text chunks are mapped into a dense embedding space, with $S, B,$ and $P$ indexed as attributes.
\end{enumerate}

\noindent \textbf{Information Transmission Mechanism.} The framework enables efficient cross-module transmission via a metadata-driven protocol. During retrieval, storage path $S$ allows direct raw multimodal data transmission to the VLM. Meanwhile, leveraging $B$ and $P$, the system fetches adjacent context from the original document, effectively bridging the ``Information Gap'' between isolated fragments and their source context.

\subsection{Autothink Router and Functional Decoupling Architecture}
\label{sec:router_fda}

This section details the design motivation and technical implementation of the Query Complexity Router (QCR) and the Decomposition of Perception and Reasoning (DPR) strategy.

\subsubsection{Query Complexity Router (QCR)}
The fidelity of RAG systems is fundamentally constrained by the coupling between query intents and corresponding contexts. Given the sequential nature of the pipeline, query comprehension serves as a critical bottleneck. When encountering queries of unknown complexity, conventional end-to-end approaches often necessitate large-scale models to parse implicit meanings or nested issues, incurring high costs. The QCR addresses this by directly assessing query complexity to influence subsequent retrieval and processing paths, thereby indirectly bolstering model accuracy. We employ an SLM to determine whether a query warrants decomposition into sub-questions, routing it to optimal execution paths at minimal computational cost.

\textbf{Technical Implementation}: The router extracts three categories of features from $Q$:
\begin{enumerate}[label=(\alph*), leftmargin=*]
    \item \textit{Semantic Feature}: Captures core intent (e.g., factual retrieval, comparative reasoning, causal analysis);
    \item \textit{Element Feature}: Counts entities $|\mathcal{E}(Q)|$, visual references $|\mathcal{M}(Q)|$, and constraints;
    \item \textit{Dependency Feature}: Identifies cross-chunk/multi-step reasoning needs ($\text{ctx}(Q)$).
\end{enumerate}
Based on these, QCR assigns a complexity label $c(Q)$ and generates routing instructions $I_p$.Concretely, queries relying on localized evidence are regarded as \textit{Simple}, whereas queries requiring multi-entity, multimodal aggregation and multi-step reasoning are regarded as \textit{Complex}.
\input{table/table4}
\input{table/table1}
\input{table/table2}

\subsubsection{Functional Decoupling via DPR}
Recent studies indicate that VLMs exhibit significantly lower capabilities in logical reasoning compared to LLMs, particularly in document question answering tasks involving unanswerable queries. As reported in MMGraphRAG \cite{mmgraphrag2024}, LLM-based systems consistently outperform multimodal models on the DocBench and MMLongBench benchmarks. Notably, VLM-based models tend to produce hallucinated answers when the required evidence is absent (as shown in Table~\ref{tab:combined_results}). In long-document multimodal QA, visual perception, cross-page evidence alignment, and reasoning are tightly coupled within a single generation, making the model prone to cascading errors: weakly aligned evidence causes overcommitment to unsupported answers instead of reliable abstention.Conversely, VLMs demonstrate superior performance in descriptive tasks post-perception. Consequently, we propose the Decoupled Perception and Reasoning (DPR) strategy, partitioning the task into two specialized sub-processes:

\begin{enumerate}
    \item \textbf{Visual Perception (Small-scale VLM)}: A lightweight VLM (e.g., Qwen2.5-VL-3B) serves as a visual interpreter, generating a structured description $T_v$ in a training-free, zero-shot mode:
    \begin{equation}
    T_v = \text{VLM}_{\text{small}}(v, P)
    \end{equation}
    
    \item \textbf{Logical Reasoning (LLM)}: Multimodal DocQA is transformed into text-only reasoning. A unified context $C = \{ T_v \} \cup \mathcal{R}(Q, \mathcal{D}) \cup \{ \text{Intent}(Q) \}$ is constructed. The LLM then synthesizes the answer $A$ guided by path-specific instructions:
    \begin{itemize}
        \item \textit{Simple}: ``Extract target row/column value from table descriptions.''
        \item \textit{Moderate}: ``Calculate/compare data using table descriptions (show steps).''
        \item \textit{Complex}: ``Integrate multi-table/text data for step-by-step analysis.''
    \end{itemize}
    \begin{equation}
    A = \text{LLM}(C, \text{Instruction}(c(Q)))
    \end{equation}
\end{enumerate}

By decoupling perception from reasoning, the LLM performs rigorous deduction over precise textualized visual evidence, effectively bridging the reasoning gap of standalone multimodal models.

%% file: table/table4.tex
\begin{table}[H] 
\centering
\small 
\renewcommand{\arraystretch}{1.1} 

\resizebox{\columnwidth}{!}{%
\begin{tabular}{lccccccccc}
\toprule
\multirow{2}{*}{\textbf{Doc}} & \multicolumn{5}{c}{\textbf{Types}} & \multicolumn{3}{c}{\textbf{Domains}} & \multirow{2}{*}{\textbf{All}} \\
\cmidrule(lr){2-6} \cmidrule(lr){7-9}
& Aca. & Fin. & Gov. & Law. & News & Txt. & Mm. & Una. & \\
\midrule
LLM  & 41.3 & 16.3 & 50.7 & 49.7 & 77.3 & 53.9 & 20.1 & 75.8 & 44.8 \\
MLLM & 19.8 & 16.3 & 28.4 & 31.4 & 46.5 & 35.7 & 15.9 & 39.5 & 27.7 \\
\bottomrule
\end{tabular}
}

\vspace{3mm} 

\resizebox{\columnwidth}{!}{%
\begin{tabular}{lcccccccc} 
\toprule
\multirow{2}{*}{\textbf{MM}} & \multicolumn{3}{c}{\textbf{Locations}} & \multicolumn{4}{c}{\textbf{Formats}} & \textbf{Overall} \\
\cmidrule(lr){2-4} \cmidrule(lr){5-8}
& Sin. & Mul. & Una. & C.T. & Txt. & Lay. & Fig. & \textbf{All} \\
\midrule
LLM  & 22.5 & 20.0 & 53.2 & 14.7 & 33.3 & 23.5 & 16.1 & 27.8 \\
MLLM & 13.3 & 7.9 & 13.9 & 8.4 & 10.7 & 11.8 & 11.4 & 11.6 \\
\bottomrule
\end{tabular}
}

\caption{Comparative experimental results of VLM and LLM on DocBench and MMLongBench benchmarks.}
\label{tab:combined_results}
\end{table}

%% file: table/table1.tex
\begin{table*}[t]
\centering
\renewcommand{\arraystretch}{1.2}
\setlength{\tabcolsep}{2.6mm}

\begin{tabular}{lccccccccc}
\toprule

\multirow{2}{*}{\textbf{Method}} & \multicolumn{5}{c}{\textbf{Domains}} & \multicolumn{3}{c}{\textbf{Types}} & \multirow{2}{*}{\textbf{Overall}} \\
\cmidrule(lr){2-6} \cmidrule(lr){7-9}
& Aca. & Fin. & Gov. & Law. & News. & Txt. & Mm. & Una. & \\

\midrule
AutothinkRAG & \textbf{80.16} & \textbf{75.40} & \textbf{85.71} & \textbf{87.09} & \textbf{92.50} & \textbf{86.43} & \textbf{77.31} & \textbf{81.25} & \textbf{82.13} \\
Nothink & 77.34 & 73.46 & 80.59 & 85.29 & 88.78 & 84.08 & 76.12 & 77.84 & 79.37 \\
VLM only & 77.03 & 74.00 & 81.54 & 84.39 & 89.26 & 85.32 & 75.09 & 76.42 & 79.47 \\
baseline & 76.34 & 72.93 & 79.77 & 84.07 & 84.52 & 84.65 & 74.47 & 75.83 & 78.02 \\
RAG-Anything & 74.90 & 70.68 & 77.41 & 83.06 & 81.67 & 84.21 & 76.36 & 52.80 & 75.47 \\
\bottomrule
\end{tabular}

\caption{Accuracy (\%) on DocBench Dataset. Domain categories include Academia (Aca.), Finance
(Fin.), Government (Gov.), Legal Documents (Law), and News Articles (News). Document types are
categorized as Text-only (Txt.), Multimodal (Mm.), and Unanswerable queries (Una.).}

\label{tab:docbench}

\end{table*}

%% file: table/table2.tex
\begin{table*}[t]
\centering
\renewcommand{\arraystretch}{1.2}
\setlength{\tabcolsep}{3.1mm}

\begin{tabular}{lcccccccc}
\toprule
\multirow{2}{*}{\textbf{Method}} & \multicolumn{7}{c}{Domains} & \multirow{2}{*}{\textbf{Overall}} \\
\cmidrule(lr){2-8}  
 & Res. & Tut. & Acad. & Guid. & Broch. & Admin. & Fin. & \\
\midrule
AutothinkRAG & \textbf{44.86} & \textbf{52.20} & \textbf{52.26} & \textbf{52.34} & \textbf{51.00} & \textbf{62.19} & \textbf{56.14} & \textbf{51.29} \\
Nothink & 43.15 & 48.55 & 43.21 & 52.25 & 46.00 & 53.08 & 50.42 & 47.21 \\
VLM (direct) & 41.43 & 49.27 & 43.43 & 52.25 & 41.00 & 54.32 & 48.72 & 45.55 \\
VLM (unguided) & 42.07 & 51.12 & 45.24 & 52.34 & 43.00 & 55.78 & 50.13 & 47.21 \\
baseline & 40.41 & 47.82 & 42.42 & 50.96 & 42.00 & 51.85 & 46.15 & 44.86 \\
8B & 42.80 & 44.12 & 43.21 & 52.34 & 46.00 & 50.61 & 50.42 & 46.09 \\
RAG-Anything & 41.09 & 41.30 & 44.22 & 45.16 & 40.00 & 54.32 & 52.13 & 44.36 \\
\bottomrule
\end{tabular}

\caption{Accuracy (\%) on MMLongBench across different domains and overall performance. Domain categories include Research Reports/Introductions (Res.), Tutorials/Workshops (Tut.), Academic Papers (Acad.), Guidebooks (Guid.), Brochures (Broch.), Administration/Industry Files (Admin.), and Financial Reports (Fin.).}
\label{tab:mmlongbench}
\end{table*}

%% file: sections/experiment.tex
\section{Experiments}
\label{sec:experiments}
\input{image/graphs}
In this section, we empirically validate the effectiveness of AutothinkRAG. Our goal is to address two primary research questions: (1) Compared to static retrieval, can the Autothink router effectively reduce hallucinations and handle complex query intents? (2) Does the Decoupled Perception-Reasoning (DPR) architecture demonstrate superior performance in reasoning tasks involving multimodal documents of varying lengths?

\subsection{Experimental Setup}

\paragraph{Benchmarks.}
We evaluate our framework on two comprehensive multimodal benchmarks:
\begin{itemize}
    \item \textbf{DocBench} : A diverse dataset covering five domains (Academic, Financial, Government, Law, News) and three query types (Text-only, Multimodal, and Unanswerable). The "Unanswerable" category is particularly critical for assessing model hallucination and refusal capabilities.
    \item \textbf{MMLongBench} : A benchmark designed to test long-context understanding across varied document types (e.g., Research Papers, Manuals, Financial Reports), requiring the model to synthesize information across multiple pages and modalities.
\end{itemize}

\subsection{Implementation Details}
In our experimental evaluation of AutothinkRAG, we implement a multi-model collaborative framework characterized by a decoupled architecture. Specifically, we employ \texttt{Qwen2.5-VL-3B} as the Autothink Router to facilitate global task orchestration and intent scheduling. For the knowledge construction phase, \texttt{Qwen3-8B} (non-reasoning version) is utilized to perform robust entity and relation extraction. To optimize retrieval precision, we adopt \texttt{Qwen3-embedding-8B} (with 1024-dimensional embeddings) for dense vector representation, followed by \texttt{bge-reranker-v2-m3} for fine-grained reranking. To realize the Decoupled Perception-Reasoning (DPR) paradigm, we deploy \texttt{Qwen2.5-VL-7B} as the Perception Expert, which distills complex visual information into dense textual descriptions. Subsequently, \texttt{Qwen3-32B-Instruct} serves as the Reasoning Expert to execute high-level logical deductions based on the integrated context. Regarding hyperparameter configurations, all generative models are set with a temperature of $0.7$ and a Top-$p$ value of $0.9$. During the retrieval stage, we set the Top-$K$ for entity recall to $40$ and the number of retrieved text chunks to $20$ to ensure sufficient context coverage.

\subsection{Main Results}

\paragraph{Performance on DocBench.}
Table~\ref{tab:docbench} shows that AutothinkRAG achieves the best Overall Accuracy of \textbf{82.13\%}, outperforming the baseline (78.02\%). The largest gain appears in \textbf{Unanswerable (Una)}, where accuracy improves from 52.80\% to \textbf{81.25\%} (\textbf{+28.45\%}), indicating the effectiveness of the Autothink router in detecting insufficient evidence and reducing hallucinations. Consistent improvements on \textbf{News} (+10.83\%) and \textbf{Government} (+8.30\%) further demonstrate the advantage of Hybrid Hypergraph Retrieval in modeling complex entity relations beyond simple vector search.

\input{image/vlllmdif}

\paragraph{Performance on MMLongBench.}
Table~\ref{tab:mmlongbench} reports the results on long-context tasks. AutothinkRAG consistently outperforms the baseline across all document categories, achieving an Overall Accuracy of \textbf{51.29\%} (+6.43\%). While the baseline struggles with visual noise in long documents, our \textbf{DPR Architecture} remains effective in complex categories such as \textbf{Admin} (+10.34\%) and \textbf{Finance} (+9.99\%), bridging the ``Reasoning Gap'' by converting visual signals into textual descriptions and delegating synthesis to a specialized LLM.

To quantify the efficiency of AutothinkRAG, we analyze per-query token consumption and monetary cost (Table~\ref{tab:token_cost_tokens} and Table~\ref{tab:Monetary_cost_money}). The overall inference comprises three stages: query routing (Qwen2.5-VL-3B), visual description generation (Qwen2.5-VL-7B), and final answer synthesis (LLM). The input context of the synthesis stage accounts for the dominant cost. Without routing, pure hypergraph retrieval feeds substantial redundant context into the LLM, costing 63.0k tokens / 0.132 RMB per query. With the Autothink router, the synthesis input is significantly reduced to 51.4k tokens / 0.108 RMB, yielding savings of 18.9\% and 18.2\%, respectively. This demonstrates that the router effectively avoids unnecessary long-context reasoning by dynamically selecting retrieval paths based on query complexity.

\subsection{Baselines}
To ensure experimental rigor and fairness, we built a baseline model and employed the identical model on RAG-Anything \citep{guo2025raganything} for testing and comparison.In the baseline configuration, we bypass query decomposition and employ the VLM for direct end-to-end inference, rather than using our proposed decoupled paradigm. For the RAG-Anything benchmark, we substitute its internal components with the specific models used in our framework and maintain identical hyperparameter configurations to ensure a controlled comparison.

\subsection{Ablation Study}
To demonstrate the superiority of the proposed architecture, we conducted ablation studies by systematically removing the routing and decoupling modules, respectively. Furthermore, we performed an in-depth analysis of relevant performance metrics across varying page ranges, thereby substantiating the efficacy of these components.

\paragraph{Effectiveness of the Autothink Router.}
Our arbitrary query router is implemented using a 3B Small Language Model (SLM). To evaluate its contribution, we conduct an ablation study where the router is removed while the subsequent decoupled reasoning module is retained. We report the distribution of hypergraphs across varying page ranges (Figure~\ref{fig:graph}) and the accuracy across diverse scenarios (Table~\ref{tab:mmlongbench}). The results demonstrate that the SLM router significantly reduces the proportion of hypergraphs, an effect that becomes more pronounced as the document length increases. This indicates that the router effectively mitigates information construction complexity during initial query processing. Consequently, it yields accuracy gains by simplifying complex reasoning tasks, such as multi-hop question answering.
\input{table/compare}
\input{table/compare1}
\paragraph{Effectiveness of the Decomposition of Perception and Reasoning.}
Our decoupling module transforms the direct reasoning of the VLM into a hybrid VLM+LLM indirect reasoning paradigm. To evaluate its efficacy, we conduct an ablation study by replacing this module with a standalone LLM reasoning component while keeping the upstream router constant. We evaluate the accuracy across various page ranges, as illustrated in Figure ~\ref{fig:vllm}. The results indicate that both direct VLM reasoning and unguided VLM decoupling for visual perception in long-document settings substantially underperform the proposed full decoupling framework, with the performance gap becoming more pronounced as the document length increases.

%% file: image/graphs.tex
\begin{figure*}[t]
    \centering
    \includegraphics[width=\textwidth, keepaspectratio]{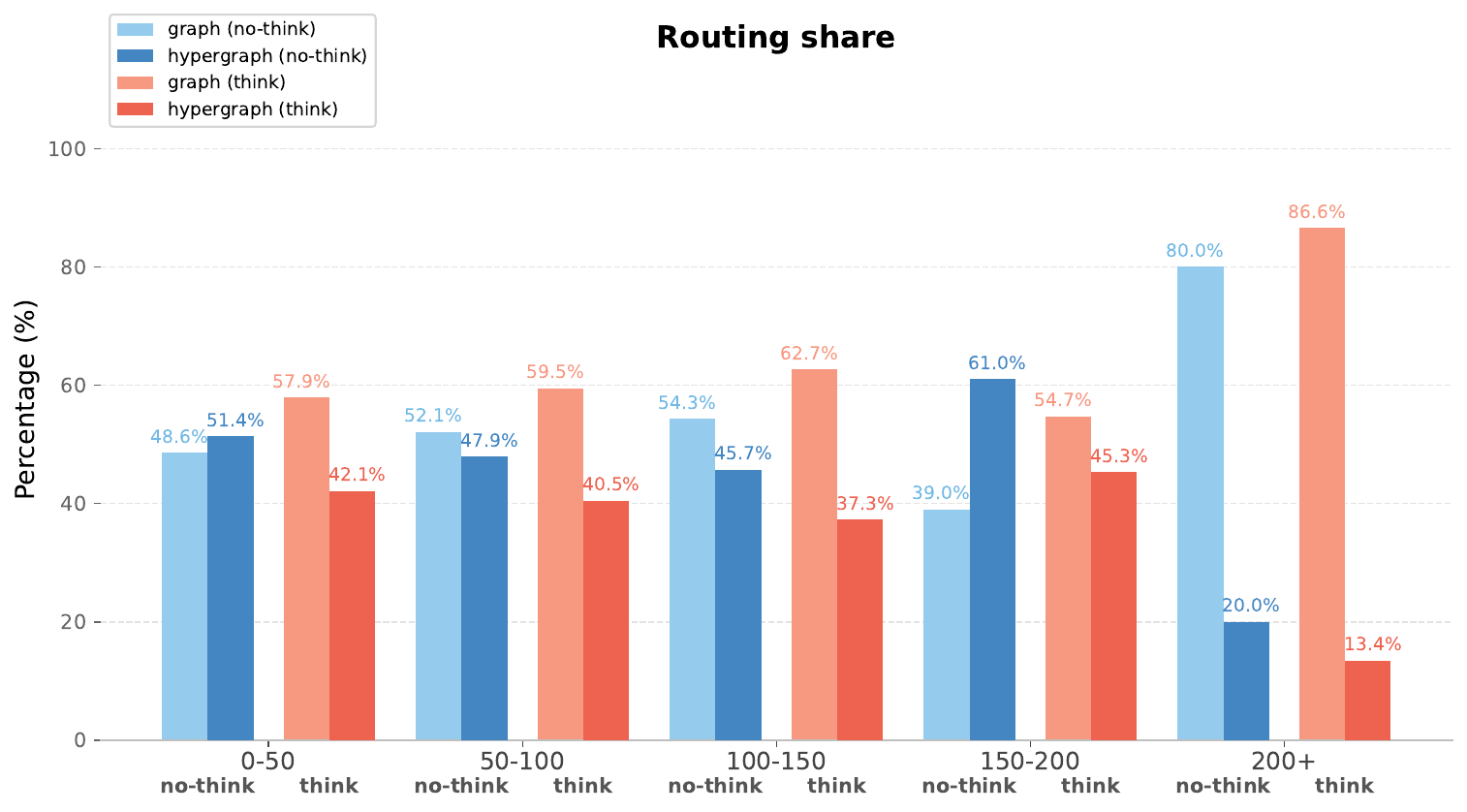}
    \caption{
    Comparison of routing strategy distributions across different document lengths. This figure illustrates the selection proportions of graphs and hypergraphs within length intervals under Think and No-Think modes.
    }
    \label{fig:graph}
\end{figure*}

%% file: image/vlllmdif.tex
\begin{figure}[t]
    \centering
    \hspace*{-4mm}\includegraphics[width=\columnwidth]{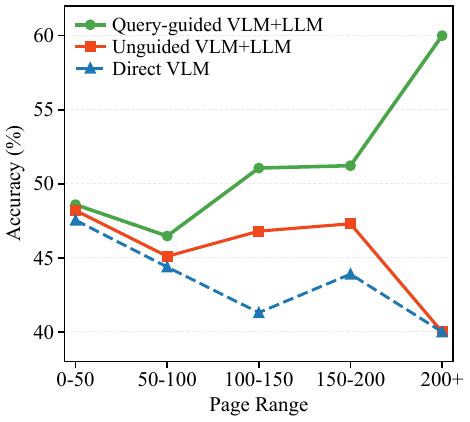}
    \caption{The approach of converting VLM outputs to LLM leads to a marked boost in accuracy when processing multi-page long documents.}
    \label{fig:vllm}
\end{figure}

%% file: table/compare.tex
\begin{table}[H]
\centering
\small
\begin{tabular}{@{\hspace{0pt}}l@{\hspace{-12pt}}c@{\hspace{12pt}}c@{\hspace{10pt}}c@{\hspace{2pt}}}
\toprule
\textbf{Pipeline} & \textbf{Image Description} & \textbf{Summary} & \textbf{Total} \\
\midrule
Hypergraph & 2.7k & 60.3k & \textbf{63.0k} \\
Routing-Enhanced & 2.7k & 48.3k & \textbf{51.4k} \\
\midrule
Savings (Abs.) & - & 12.0k & \textbf{11.6k} \\
Savings (Rel.) & - & 19.9\% & \textbf{18.9\%} \\
\bottomrule
\end{tabular}
\caption{Token comparison across stages.}
\label{tab:token_cost_tokens}
\end{table}

%% file: table/compare1.tex
\begin{table}[H]
\centering
\small
\setlength{\tabcolsep}{4pt}
\begin{tabular}{@{\hspace{0pt}}l@{\hspace{-12pt}}c@{\hspace{12pt}}c@{\hspace{10pt}}c@{\hspace{2pt}}}
\toprule
\textbf{Pipeline} & \textbf{Image Description} & \textbf{Summary} & \textbf{Total} \\
\midrule
Hypergraph & 0.0084 & 0.1239 & \textbf{0.132} \\
Routing-Enhanced & 0.0084 & 0.0984 & \textbf{0.108} \\
\midrule
Savings (Abs.) & - & 0.0255 & \textbf{0.024} \\
Savings (Rel.) & - & 20.6\% & \textbf{18.2\%} \\
\bottomrule
\end{tabular}
\caption{Monetary cost comparison across stages.}
\label{tab:Monetary_cost_money}
\end{table}

%% file: sections/conclusion.tex
\section{Conclusion}
We present AutothinkRAG, a multi-model collaborative framework that combines a Query Complexity Router with a functionally-decoupled pipeline to tackle long-context and information-overload challenges in information-intensive DocQA. By leveraging a lightweight VLM as a high-fidelity visual interpreter and an LLM for logical deduction, AutothinkRAG sets new state-of-the-art results on DocBench and MMLongBench while significantly cutting inference costs. Extensive experiments and ablations confirm the effectiveness of both the router and the decoupled architecture.

\section*{Limitations}

While AutothinkRAG achieves significant performance gains over existing multimodal GraphRAG architectures using models with fewer parameters, it still relies on a sequential document parsing and embedding pipeline, which limits the overall processing speed. Our future work will be dedicated to achieving the efficient coupling of document parsing and information encoding to improve the overall processing speed and accuracy.

\section*{Ethical Considerations}
All our experiments are conducted using widely adopted and recognized open-source datasets curated for academic research purposes. Consequently, these datasets do not contain any inappropriate or sensitive information. While our framework achieves state-of-the-art performance, it is important to acknowledge that our model also builds upon community-driven open-source models and is not flawless. Therefore, users should exercise caution and not rely exclusively on the generated results. In real-world applications, human intervention is recommended to ensure accuracy.

%% file: acl_latex.bbl
\begin{thebibliography}{45}
\providecommand{\natexlab}[1]{#1}

\bibitem[{Chen et~al.(2025{\natexlab{a}})Chen, Guo, Yang, Chen, Chen, Liu, Shi, and Yang}]{liu2024pathrag}
Boyu Chen, Zirui Guo, Zidan Yang, Yuluo Chen, Junze Chen, Zhenghao Liu, Chuan Shi, and Cheng Yang. 2025{\natexlab{a}}.
\newblock \href {https://arxiv.org/abs/2502.14902} {Pathrag: Pruning graph-based retrieval augmented generation with relational paths}.
\newblock \emph{arXiv preprint arXiv:2502.14902}.

\bibitem[{Chen et~al.(2026)Chen, Yang, Li, Zhang, Hu, and Zhang}]{beyondchunking2023}
Huiyao Chen, Yi~Yang, Yinghui Li, Meishan Zhang, Baotian Hu, and Min Zhang. 2026.
\newblock \href {https://arxiv.org/abs/2506.06313} {Beyond chunking: Discourse-aware hierarchical retrieval for long document question answering}.

\bibitem[{Chen et~al.(2025{\natexlab{b}})Chen, Xu, Fei, Feng, and Elhoseiny}]{chen2025haystacks}
Jun Chen, Dannong Xu, Junjie Fei, Chun-Mei Feng, and Mohamed Elhoseiny. 2025{\natexlab{b}}.
\newblock \href {https://arxiv.org/abs/2411.16740} {Document haystacks: Vision-language reasoning over piles of 1000+ documents}.
\newblock In \emph{Proceedings of the Computer Vision and Pattern Recognition Conference}, pages 24817--24826.

\bibitem[{Choi et~al.(2017)Choi, Hewlett, Uszkoreit, Polosukhin, Lacoste, and Berant}]{choi2017coarse}
Eunsol Choi, Daniel Hewlett, Jakob Uszkoreit, Illia Polosukhin, Alexandre Lacoste, and Jonathan Berant. 2017.
\newblock \href {https://aclanthology.org/P17-1020/} {Coarse-to-fine question answering for long documents}.
\newblock In \emph{Proceedings of the 55th Annual Meeting of the Association for Computational Linguistics (Volume 1: Long Papers)}, pages 209--220.

\bibitem[{Edge et~al.(2024{\natexlab{a}})Edge, Trinh, Cheng, Bradley, Chao, Mody, Truitt, Metropolitansky, Ness, and Larson}]{edge2024graphrag}
Darren Edge, Ha~Trinh, Newman Cheng, Joshua Bradley, Alex Chao, Apurva Mody, Steven Truitt, Dasha Metropolitansky, Robert~Osazuwa Ness, and Jonathan Larson. 2024{\natexlab{a}}.
\newblock \href {https://arxiv.org/abs/2404.16130} {From local to global: A graph rag approach to query-focused summarization}.
\newblock \emph{arXiv preprint arXiv:2404.16130}.

\bibitem[{Edge et~al.(2024{\natexlab{b}})Edge, Trinh, Cheng, Bradley, Chao, Mody, Truitt, Metropolitansky, Ness, and Larson}]{chen2024localglobal}
Darren Edge, Ha~Trinh, Newman Cheng, Joshua Bradley, Alex Chao, Apurva Mody, Steven Truitt, Dasha Metropolitansky, Robert~Osazuwa Ness, and Jonathan Larson. 2024{\natexlab{b}}.
\newblock \href {https://arxiv.org/abs/2404.16130} {From local to global: A graph rag approach to query-focused summarization}.
\newblock \emph{arXiv preprint arXiv:2404.16130}.

\bibitem[{Fang et~al.(2024{\natexlab{a}})Fang, Meng, and Macdonald}]{li2024reano}
Jinyuan Fang, Zaiqiao Meng, and Craig Macdonald. 2024{\natexlab{a}}.
\newblock \href {https://aclanthology.org/2024.acl-long.115/} {Reano: Optimising retrieval-augmented reader models through knowledge graph generation}.
\newblock In \emph{Proceedings of the 62nd Annual Meeting of the Association for Computational Linguistics (Volume 1: Long Papers)}, pages 2094--2112.

\bibitem[{Fang et~al.(2024{\natexlab{b}})Fang, Meng, and Macdonald}]{trivedi2023trace}
Jinyuan Fang, Zaiqiao Meng, and Craig Macdonald. 2024{\natexlab{b}}.
\newblock \href {https://aclanthology.org/2024.findings-emnlp.496/} {Trace the evidence: Constructing knowledge-grounded reasoning chains for retrieval-augmented generation}.
\newblock \emph{arXiv preprint arXiv:2406.11460}.

\bibitem[{Faysse et~al.(2024)Faysse, Sibille, Wu, Omrani, Viaud, Hudelot, and Colombo}]{delatorre2024colpali}
Manuel Faysse, Hugues Sibille, Tony Wu, Bilel Omrani, Gautier Viaud, C{\'e}line Hudelot, and Pierre Colombo. 2024.
\newblock \href {https://openreview.net/forum?id=ogjBpZ8uSi} {Colpali: Efficient document retrieval with vision language models}.
\newblock \emph{arXiv preprint arXiv:2407.01449}.

\bibitem[{Gu et~al.(2023)Gu, Gao, and Hahnloser}]{memsumdqa2022}
Nianlong Gu, Yingqiang Gao, and Richard~HR Hahnloser. 2023.
\newblock \href {https://arxiv.org/abs/2310.06436} {Memsum-dqa: Adapting an efficient long document extractive summarizer for document question answering}.
\newblock \emph{arXiv preprint arXiv:2310.06436}.

\bibitem[{Guo et~al.(2025)Guo, Ren, Xu, Zhang, and Huang}]{guo2025raganything}
Zirui Guo, Xubin Ren, Lingrui Xu, Jiahao Zhang, and Chao Huang. 2025.
\newblock \href {https://arxiv.org/abs/2510.12323} {Rag-anything: All-in-one rag framework}.
\newblock \emph{arXiv preprint arXiv:2510.12323}.

\bibitem[{Guo et~al.(2024)Guo, Xia, Yu, Ao, and Huang}]{guo2024lightrag}
Zirui Guo, Lianghao Xia, Yanhua Yu, Tu~Ao, and Chao Huang. 2024.
\newblock \href {https://arxiv.org/abs/2410.05779} {Lightrag: Simple and fast retrieval-augmented generation}.
\newblock \emph{arXiv preprint arXiv:2410.05779}.

\bibitem[{He et~al.(2024)He, Tian, Sun, Chawla, Laurent, LeCun, Bresson, and Hooi}]{sun2024gretriever}
Xiaoxin He, Yijun Tian, Yifei Sun, Nitesh Chawla, Thomas Laurent, Yann LeCun, Xavier Bresson, and Bryan Hooi. 2024.
\newblock \href {https://proceedings.neurips.cc/paper_files/paper/2024/file/efaf1c9726648c8ba363a5c927440529-Paper-Conference.pdf} {G-retriever: Retrieval-augmented generation for textual graph understanding and question answering}.
\newblock \emph{Advances in Neural Information Processing Systems}, 37:132876--132907.

\bibitem[{Hu et~al.(2023)Hu, Iscen, Sun, Wang, Chang, Sun, Schmid, Ross, and Fathi}]{hu2023reveal}
Ziniu Hu, Ahmet Iscen, Chen Sun, Zirui Wang, Kai-Wei Chang, Yizhou Sun, Cordelia Schmid, David~A Ross, and Alireza Fathi. 2023.
\newblock \href {https://arxiv.org/pdf/2212.05221} {Reveal: Retrieval-augmented visual-language pre-training with multi-source multimodal knowledge memory}.
\newblock In \emph{Proceedings of the IEEE/CVF conference on computer vision and pattern recognition}, pages 23369--23379.

\bibitem[{Iscen et~al.(2023)Iscen, Caron, Fathi, and Schmid}]{retrievalcontrastivevlm2023}
Ahmet Iscen, Mathilde Caron, Alireza Fathi, and Cordelia Schmid. 2023.
\newblock \href {https://openreview.net/forum?id=b2UlHeyyC0} {Retrieval-enhanced contrastive vision-text models}.
\newblock \emph{arXiv preprint arXiv:2306.07196}.

\bibitem[{Jiang et~al.(2024)Jiang, Qiu, Xu, Zhu, Zhang, Fang, Xu, Zhao, and Wang}]{ma2024ragraph}
Xinke Jiang, Rihong Qiu, Yongxin Xu, Yichen Zhu, Ruizhe Zhang, Yuchen Fang, Chu Xu, Junfeng Zhao, and Yasha Wang. 2024.
\newblock \href {https://arxiv.org/abs/2410.23855} {Ragraph: A general retrieval-augmented graph learning framework}.
\newblock \emph{Advances in Neural Information Processing Systems}, 37:29948--29985.

\bibitem[{Jiang et~al.(2023)Jiang, Xu, Gao, Sun, Liu, Dwivedi-Yu, Yang, Callan, and Neubig}]{activerag2023}
Zhengbao Jiang, Frank~F Xu, Luyu Gao, Zhiqing Sun, Qian Liu, Jane Dwivedi-Yu, Yiming Yang, Jamie Callan, and Graham Neubig. 2023.
\newblock \href {https://aclanthology.org/2023.emnlp-main.495/} {Active retrieval augmented generation}.
\newblock In \emph{Proceedings of the 2023 Conference on Empirical Methods in Natural Language Processing}, pages 7969--7992.

\bibitem[{Lee and Whang()}]{zhou2024structure}
Jaejun Lee and Joyce~Jiyoung Whang.
\newblock \href {https://openreview.net/attachment?id=2tH2vexW1Z&name=pdf} {Structure is all you need: Structural representation learning on hyper-relational knowledge graphs}.
\newblock In \emph{Forty-second International Conference on Machine Learning}.

\bibitem[{Lee et~al.(2025)Lee, Kim, Kang, Bang, Song, and Lee}]{li2023ratta}
Youngjun Lee, Doyoung Kim, Junhyeok Kang, Jihwan Bang, Hwanjun Song, and Jae-Gil Lee. 2025.
\newblock \href {https://openreview.net/forum?id=V3zobHnS61} {Ra-tta: Retrieval-augmented test-time adaptation for vision-language models}.
\newblock In \emph{The Thirteenth International Conference on Learning Representations}.

\bibitem[{Liu et~al.(2025)Liu, Wang, Chen, Li, Xiong, Yu, and Zhang}]{wang2024hoprag}
Hao Liu, Zhengren Wang, Xi~Chen, Zhiyu Li, Feiyu Xiong, Qinhan Yu, and Wentao Zhang. 2025.
\newblock \href {https://arxiv.org/abs/2502.12442} {Hoprag: Multi-hop reasoning for logic-aware retrieval-augmented generation}.
\newblock \emph{arXiv preprint arXiv:2502.12442}.

\bibitem[{Luo et~al.(2025)Luo, Chen, Zheng, Wu, Guo, Lin, Feng, Kuang, Song, Zhu et~al.}]{luo2025hypergraphrag}
Haoran Luo, Guanting Chen, Yandan Zheng, Xiaobao Wu, Yikai Guo, Qika Lin, Yu~Feng, Zemin Kuang, Meina Song, Yifan Zhu, and 1 others. 2025.
\newblock \href {https://arxiv.org/abs/2503.21322} {Hypergraphrag: Retrieval-augmented generation via hypergraph-structured knowledge representation}.
\newblock \emph{arXiv preprint arXiv:2503.21322}.

\bibitem[{Mass et~al.(2024)Mass, Carmeli, Yehudai, Toledo, and Mills}]{ma2024virtualdocs}
Yosi Mass, Boaz Carmeli, Asaf Yehudai, Assaf Toledo, and Nathaniel Mills. 2024.
\newblock \href {https://aclanthology.org/2024.findings-emnlp.757/} {More bang for your context: Virtual documents for question answering over long documents}.
\newblock In \emph{Findings of the Association for Computational Linguistics: EMNLP 2024}, pages 12936--12942.

\bibitem[{Mavromatis and Karypis(2024)}]{zhang2024gnnrag}
Costas Mavromatis and George Karypis. 2024.
\newblock \href {https://aclanthology.org/2025.findings-acl.856/} {Gnn-rag: Graph neural retrieval for large language model reasoning}.
\newblock \emph{arXiv preprint arXiv:2405.20139}.

\bibitem[{Nie et~al.(2022)Nie, Huang, Wei, and Mao}]{compressivegraphselector2022}
Yuxiang Nie, He-Yan Huang, Wei Wei, and Xian-Ling Mao. 2022.
\newblock \href {https://arxiv.org/abs/2210.05499} {Capturing global structural information in long document question answering with compressive graph selector network}.
\newblock In \emph{Proceedings of the 2022 Conference on Empirical Methods in Natural Language Processing}, pages 5036--5047.

\bibitem[{Ren et~al.(2025)}]{shi2024videorag}
X.~Ren and 1 others. 2025.
\newblock \href {https://arxiv.org/abs/2502.01549} {Videorag: Retrieval-augmented generation with extreme long-context videos}.
\newblock \emph{arXiv preprint arXiv:2502.01549}.

\bibitem[{Saad-Falcon et~al.(2024)Saad-Falcon, Barrow, Siu, Nenkova, Yoon, Rossi, and Dernoncourt}]{pdftriage2023}
Jon Saad-Falcon, Joe Barrow, Alexa Siu, Ani Nenkova, Seunghyun Yoon, Ryan~A Rossi, and Franck Dernoncourt. 2024.
\newblock \href {https://aclanthology.org/2024.emnlp-industry.13/} {Pdftriage: Question answering over long, structured documents}.
\newblock In \emph{Proceedings of the 2024 Conference on Empirical Methods in Natural Language Processing: Industry Track}, pages 153--169.

\bibitem[{Shao et~al.(2023)Shao, Gong, Shen, Huang, Duan, and Chen}]{iterativerag2023}
Zhihong Shao, Yeyun Gong, Yelong Shen, Minlie Huang, Nan Duan, and Weizhu Chen. 2023.
\newblock \href {https://arxiv.org/abs/2305.15294} {Enhancing retrieval-augmented large language models with iterative retrieval-generation synergy}.
\newblock \emph{arXiv preprint arXiv:2305.15294}.

\bibitem[{Shi et~al.(2024)Shi, Zhang, Sun, Gao, Ren, Chen, and Ren}]{zhao2023generate}
Zhengliang Shi, Shuo Zhang, Weiwei Sun, Shen Gao, Pengjie Ren, Zhumin Chen, and Zhaochun Ren. 2024.
\newblock \href {https://aclanthology.org/2024.acl-long.397/} {Generate-then-ground in retrieval-augmented generation for multi-hop question answering}.
\newblock In \emph{Proceedings of the 62nd Annual Meeting of the Association for Computational Linguistics (Volume 1: Long Papers)}, pages 7339--7353.

\bibitem[{Sun et~al.(2025)Sun, Zhao, Han, and Xiong}]{domainrag2023}
Liwen Sun, James~Jialun Zhao, Wenjing Han, and Chenyan Xiong. 2025.
\newblock \href {https://aclanthology.org/2025.naacl-long.28/} {Fact-aware multimodal retrieval augmentation for accurate medical radiology report generation}.
\newblock In \emph{Proceedings of the 2025 Conference of the Nations of the Americas Chapter of the Association for Computational Linguistics: Human Language Technologies (Volume 1: Long Papers)}, pages 643--655.

\bibitem[{Suri et~al.(2025)}]{zhao2025visdom}
Manan Suri and 1 others. 2025.
\newblock \href {https://aclanthology.org/2025.naacl-long.310/} {Visdom: Multi-document qa with visually rich elements using multimodal rag}.
\newblock In \emph{NAACL 2025}.

\bibitem[{Tanaka et~al.(2025)Tanaka, Iki, Hasegawa, Nishida, Saito, and Suzuki}]{vdocgrag2023}
Ryota Tanaka, Taichi Iki, Taku Hasegawa, Kyosuke Nishida, Kuniko Saito, and Jun Suzuki. 2025.
\newblock \href {https://openaccess.thecvf.com/content/CVPR2025/papers/Tanaka_VDocRAG_Retrieval-Augmented_Generation_over_Visually-Rich_Documents_CVPR_2025_paper.pdf} {Vdocrag: Retrieval-augmented generation over visually-rich documents}.
\newblock In \emph{Proceedings of the Computer Vision and Pattern Recognition Conference}, pages 24827--24837.

\bibitem[{Wan and Yu(2025)}]{mmgraphrag2024}
Xueyao Wan and Hang Yu. 2025.
\newblock \href {https://arxiv.org/abs/2507.20804} {Mmgraphrag: Bridging vision and language with interpretable multimodal knowledge graphs}.
\newblock \emph{arXiv preprint arXiv:2507.20804}.

\bibitem[{Wang et~al.(2025{\natexlab{a}})Wang, Fang, Zhou, Liu, and Ma}]{wu2024archrag}
Shu Wang, Yixiang Fang, Yingli Zhou, Xilin Liu, and Yuchi Ma. 2025{\natexlab{a}}.
\newblock \href {https://arxiv.org/abs/2502.09891} {Archrag: Attributed community-based hierarchical retrieval-augmented generation}.
\newblock \emph{arXiv preprint arXiv:2502.09891}.

\bibitem[{Wang et~al.(2025{\natexlab{b}})Wang, Zhou, and Fang}]{bookrag2023}
Shu Wang, Yingli Zhou, and Yixiang Fang. 2025{\natexlab{b}}.
\newblock \href {https://arxiv.org/abs/2512.03413} {Bookrag: A hierarchical structure-aware index-based approach for retrieval-augmented generation on complex documents}.

\bibitem[{Wang et~al.(2024)Wang, Ren, Li, Zhao, Liu, and Wen}]{li2024rear}
Yuhao Wang, Ruiyang Ren, Junyi Li, Wayne~Xin Zhao, Jing Liu, and Ji-Rong Wen. 2024.
\newblock \href {https://aclanthology.org/2024.emnlp-main.321/} {Rear: A relevance-aware retrieval-augmented framework for open-domain question answering}.

\bibitem[{Wang et~al.(2025{\natexlab{c}})Wang, Gao, Xiao, Huang, Si, Luo, Bai, Li, Duan, Lv et~al.}]{park2025segmentation}
Zhitong Wang, Cheng Gao, Chaojun Xiao, Yufei Huang, Shuzheng Si, Kangyang Luo, Yuzhuo Bai, Wenhao Li, Tangjian Duan, Chuancheng Lv, and 1 others. 2025{\natexlab{c}}.
\newblock \href {https://aclanthology.org/2025.findings-acl.422.pdf} {Document segmentation matters for retrieval-augmented generation}.
\newblock In \emph{Findings of the Association for Computational Linguistics: ACL 2025}, pages 8063--8075.

\bibitem[{Wu et~al.(2025)Wu, Xia, Yu, Chen, Harsha, Maharaj, Zhang, Bursztyn, Kim, Rossi et~al.}]{docreact2024}
Junda Wu, Yu~Xia, Tong Yu, Xiang Chen, Sai~Sree Harsha, Akash~V Maharaj, Ruiyi Zhang, Victor Bursztyn, Sungchul Kim, Ryan~A Rossi, and 1 others. 2025.
\newblock \href {https://aclanthology.org/2025.acl-short.6/} {Doc-react: Multi-page heterogeneous document question-answering}.
\newblock In \emph{Proceedings of the 63rd Annual Meeting of the Association for Computational Linguistics (Volume 2: Short Papers)}, pages 67--78.

\bibitem[{Xia et~al.(2024)Xia, Zhu, Li, Wang, Shi, Wang, Zhang, Zou, and Yao}]{zhang2023mmedrag}
Peng Xia, Kangyu Zhu, Haoran Li, Tianze Wang, Weijia Shi, Sheng Wang, Linjun Zhang, James Zou, and Huaxiu Yao. 2024.
\newblock \href {https://openreview.net/forum?id=s5epFPdIW6} {Mmed-rag: Versatile multimodal rag system for medical vision language models}.
\newblock \emph{arXiv preprint arXiv:2410.13085}.

\bibitem[{Xiong et~al.(2016)Xiong, Zhong, and Socher}]{seo2017dynamic}
Caiming Xiong, Victor Zhong, and Richard Socher. 2016.
\newblock \href {https://arxiv.org/abs/1611.01604} {Dynamic coattention networks for question answering}.
\newblock \emph{arXiv preprint arXiv:1611.01604}.

\bibitem[{Yang et~al.(2025)Yang, Simoulin, Qian, Liu, Cao, Teng, and Yang}]{docagent2024}
Dayu Yang, Antoine Simoulin, Xin Qian, Xiaoyi Liu, Yuwei Cao, Zhaopu Teng, and Grey Yang. 2025.
\newblock \href {https://arxiv.org/abs/2504.08725} {Docagent: A multi-agent system for automated code documentation generation}.
\newblock \emph{arXiv preprint arXiv:2504.08725}.

\bibitem[{Yasunaga et~al.(2022)Yasunaga, Leskovec, and Liang}]{yasunaga2022linkbert}
Michihiro Yasunaga, Jure Leskovec, and Percy Liang. 2022.
\newblock \href {https://aclanthology.org/2022.acl-long.551/} {Linkbert: Pretraining language models with document links}.
\newblock \emph{arXiv preprint arXiv:2203.15827}.

\bibitem[{Yu et~al.(2024)Yu, Tang, Xu, Cui, Ran, Yan, Liu, Wang, Han, Liu et~al.}]{singh2024visrag}
Shi Yu, Chaoyue Tang, Bokai Xu, Junbo Cui, Junhao Ran, Yukun Yan, Zhenghao Liu, Shuo Wang, Xu~Han, Zhiyuan Liu, and 1 others. 2024.
\newblock \href {https://arxiv.org/abs/2410.10594} {Visrag: Vision-based retrieval-augmented generation on multi-modality documents}.
\newblock \emph{arXiv preprint arXiv:2410.10594}.

\bibitem[{Zhang et~al.(2024)Zhang, Chong, Zhang, Li, Cheng, Yan, and Liang}]{yang2024garmentaligner}
Shiyue Zhang, Zheng Chong, Xujie Zhang, Hanhui Li, Yuhao Cheng, Yiqiang Yan, and Xiaodan Liang. 2024.
\newblock \href {https://arxiv.org/abs/2408.12352} {Garmentaligner: Text-to-garment generation via retrieval-augmented multi-level corrections}.
\newblock In \emph{European Conference on Computer Vision}, pages 148--164. Springer.

\bibitem[{Zhang et~al.(2023)Zhang, Zhang, Ren, Shi, Han, Wu, Lai, and Cao}]{iag2023}
Zhebin Zhang, Xinyu Zhang, Yuanhang Ren, Saijiang Shi, Meng Han, Yongkang Wu, Ruofei Lai, and Zhao Cao. 2023.
\newblock \href {https://aclanthology.org/2023.emnlp-main.1/} {Iag: Induction-augmented generation framework for answering reasoning questions}.

\bibitem[{Zhu et~al.(2025)Zhu, Meng, Chen, Li, Pfister, and Yoon}]{doclens2025}
Dawei Zhu, Rui Meng, Jiefeng Chen, Sujian Li, Tomas Pfister, and Jinsung Yoon. 2025.
\newblock \href {https://arxiv.org/abs/2511.11552} {Doclens: A tool-augmented multi-agent framework for long visual document understanding}.
\newblock \emph{arXiv preprint arXiv:2511.11552}.

\end{thebibliography}
